\documentclass[aip,pof,reprint]{revtex4-1}

\usepackage{graphicx}
\usepackage{amssymb}
\usepackage{amsmath}

\begin{document}

\title{Deformation of vortex patches by boundaries}
\author{A. Crosby}
\affiliation{ITG, DAMTP, University of Cambridge, Wilberforce Rd., Cambridge, CB3 0WA, UK}
\author{E.R. Johnson}
\affiliation{Department of Mathematics, University College London, Gower Street, London, WC1E 6BT, UK}
\author{P.J. Morrison}
\affiliation{Department of Physics and Institute for Fusion Studies, University of Texas, Austin, Texas 78712, US}

\begin{abstract}
The deformation of two-dimensional vortex patches in the vicinity of fluid boundaries is investigated. The presence of a boundary causes an initially circular patch of uniform vorticity to deform.  Sufficiently far away from the boundary, the deformed shape is well approximated by an ellipse. This leading order elliptical deformation is investigated via the elliptic moment model of Melander, Zabusky \& Styczek [M.~V.~Melander, N.~J.~Zabusky \& A.~S.~Styczek, J.~Fluid.~Mech., \textbf{167}, 95 (1986)]. When the boundary is straight, the centre of the elliptic patch remains at a constant distance from the boundary, and the motion is integrable.  Furthermore, since the straining flow acting on the patch is constant in time, the problem is that of an elliptic vortex patch in constant strain, which was analysed by Kida [S.~Kida, J.~Phys.~Soc.~Japan, \textbf{50}, 3517 (1981)]. For more complicated boundary shapes, such as a square corner, the motion is no longer integrable. Instead, there is an adiabatic invariant for the motion. This adiabatic invariant arises due to the separation in times scales between the relatively rapid time scale associated with the rotation of the patch and the slower time scale associated with the self-advection of the patch along the boundary.  The interaction of a vortex patch with a circular island is also considered. Without a background flow, conservation of angular impulse implies that the motion is again integrable.  The addition of an irrotational flow past the island can drive the patch towards the boundary, leading to the possibility of large deformations and breakup.
\end{abstract}

\maketitle

\section{Introduction}

Many problems in geophysical fluid dynamics feature quasi-two-dimensional flows due to the presence of rotation, stratification, or shallow water; and a ubiquitous feature of such flows is the presence of large-scale vorticity structures. Two examples can be found in the Atlantic Ocean: an instability of the Gulf Stream leads to the formation of Gulf Stream rings of relatively hot or cold water \cite{gulfstream_rings}; and the flow of water out of the Mediterranean Sea leads to the formation of Meddies, which are rotating regions of relatively warm and salty water \cite{meddies2,meddies}. In both cases, the vorticity structures can have life times of more than a year, and are important in the transport of heat and salt across the Atlantic. During their life times, these structures can interact with both coastal and submarine topography, which leads to deformation of the structures. For example, Richard~\&~Tychensky \cite{meddies} observed the interaction of Meddies with underwater seamounts, and found that such interactions could cause the Meddy to break up.

The simplest formulation within which to study such flows is the two-dimensional incompressible Euler equations. This formulation has been used by Johnson~\&~McDonald \cite{two_islands,gap_in_wall} who modelled vortex patches as point vortices in order to investigate the paths of ocean vortex patches in the vicinity of boundaries such as islands or coastal gaps. However, with the exception of a few contour dynamics calculations (e.g.\ Appendix C of Thompson \cite{thompson} and Crowdy~\&~Surana \cite{CD_complex_domains}), there has been little focus on the deformation of vorticity structures caused by the presence of boundaries. This motivates the more systematic investigation presented here.

As a model for more complex vorticity structures, we consider (initially circular) patches of uniform vorticity and utilise the elliptic model of Melander, Zabusky \& Styczek \cite{MZS}. This model has been widely used, with successful applications including the study of symmetric vortex merging both with\cite{ngan} and without\cite{vortex_merger} an imposed background shear flow. We favour the simplicity of this low-order model over the potential for increased realism from models with varying vorticity such as the elliptical model of Legras~\&~Dritschel\cite{multi_layer_elliptic}. When more complex vorticity structures have been considered, the action of an external straining flow has been found to strip off outer layers of vorticity leading to the steepening of vorticity gradients and consequently to structures that more closely resemble the uniform patches of vorticity that we consider here\cite{vortex_stripping_adiabatic}.

In Section~\ref{sec:methods}, we describe the elliptic model of Melander, Zabusky \& Styczek \cite{MZS}, and we also give a brief description of the contour dynamics method that can be used to calculate numerically the full evolution of patches of uniform vorticity. Then, in Section~\ref{sec:straight}, we apply the elliptic model to the motion of an initially circular vortex patch in the vicinity of a straight boundary. The resulting motion is integrable, and is shown to reduce to the problem of an elliptic vortex patch in constant strain, which has been analysed by Kida \cite{Kida}. Comparison with contour dynamics solutions confirms that the elliptic model is a good approximation to the motion, even when the separation of the vortex patch from the boundary is comparable to the dimensions of the vortex patch. In Section~\ref{sec:corner}, we consider the motion of a vortex patch around a corner, which is no longer integrable. The time scale associated with the rotation of the patch is found to be short compared to the time scale on which the straining flow experienced by the patch varies. This separation of time scales in a Hamiltonian system implies the existence of an adiabatic invariant, which is an (approximate) constant of the motion\cite{adiabatic}. We briefly consider the effect of adding a background straining flow in Section~\ref{sec:chaos}. This flow can trap vortex patches in the corner. We use Poincar\'{e} sections to reveal the underlying chaotic nature of the elliptic model. Then, in Section~\ref{sec:island}, we consider the deformation of a vortex patch due to a circular island. Without a background flow, the motion around an island is found to be integrable. When a background flow is added, significant deformation, and possible breakup of vortex patches is observed as they are driven towards the boundary.

\section{Methods}
\label{sec:methods}

\subsection{Elliptic model}
\label{sec:elliptic_model}

A vortex patch of uniform vorticity, in the vicinity of a boundary, will generate a flow field $\mathbf{u}(\mathbf{x})$ which can be decomposed as $\mathbf{u}(\mathbf{x}) = \mathbf{u}_{\infty}(\mathbf{x}) + \mathbf{u}_{B}(\mathbf{x})$, where $\mathbf{u}_{\infty}(\mathbf{x})$ is the flow field that the patch would generate in an unbounded geometry, and $\mathbf{u}_{B}(\mathbf{x})$ is the perturbation to that flow due to the presence of the boundary.  For an elliptical vortex patch, $\mathbf{u}_{\infty}(\mathbf{x})$ simply causes the patch to rotate while remaining the same shape \cite{Lamb}. The flow due to the boundary, $\mathbf{u}_{B}(\mathbf{x})$, can be decomposed about the centre of the vortex patch as a uniform advection, a straining flow, and higher order corrections. The uniform advection simply translates the vortex patch with no change of shape. The straining flow will change the shape of the vortex patch, but an intially elliptical vortex patch will always remain elliptical in shape \cite{Kida}. Thus, provided the higher order corrections can be safely neglected, the effect of a boundary on an initially circular vortex patch is to deform it into an elliptical shape. This observation motivates the use of an elliptic moment model to approximate the evolution of such a vortex patch.

When the boundary shape is sufficiently simple, the motion can be solved via the method of images. In this case, the motion is simply that of a collection of elliptical vortex patches.  A Hamiltonian elliptic moment model for such a collection of vortex patches was developed by Melander, Zabusky \& Styczek \cite{MZS}.  In Section~\ref{sec:elliptic_model_eqs} we will recap the equations of motion associated with this model, but first, we briefly consider how to represent an elliptic shape mathematically.

\subsection{Representing an ellipse}

For an ellipse of area $A$, two further independent quantities are needed to describe the shape.  Perhaps the simplest and most physically intuitive pairing is $(r,\theta)$, where $r \geq 1$ is the ratio of the length of the major axis of the ellipse to that of its minor axis, and $\theta$ is the angle that the major axis makes to some predetermined direction (here we will always measure angles relative to the positive $x$-axis).

A second representation can be obtained by considering the quadratic moment of an elliptic patch
\begin{equation}
\mathbf{q} \equiv \begin{pmatrix} q_{xx} & q_{xy} \\ q_{xy} & q_{yy} \end{pmatrix} \equiv \int_A (\hat{\mathbf{x}}-\mathbf{x}_c)(\hat{\mathbf{x}}-\mathbf{x}_c)\,d^2\hat{\mathbf{x}},
\end{equation}
where $\mathbf{x}_c$ is the centroid of the ellipse and the integral is taken over the area of the ellipse.  This can of course be written in terms of $r$ and $\theta$
\begin{equation}
\mathbf{q}  = \frac{A^2}{4\pi} \begin{pmatrix} r \cos^2\theta + r^{-1} \sin^2\theta & (r-r^{-1})\sin\theta\cos\theta \\ (r-r^{-1})\sin\theta\cos\theta & r \sin^2\theta + r^{-1} \cos^2\theta \end{pmatrix}.
\end{equation}
Since there are three quantities $q_{xx}$, $q_{xy}$ and $q_{yy}$ representing the shape, they can not all be independent; they satisfy the relation $\mbox{det}(\mathbf{q})= A^2/(4\pi)$.

To facilitate a compact representation of the elliptic model in the next section, we also define a matrix $\mathbf{p}$, closely related to $\mathbf{q}$, as
\begin{equation}
\mathbf{p} \equiv \begin{pmatrix} q_{xy} & (q_{yy}-q_{xx})/2 \\ (q_{yy}-q_{xx})/2 & -q_{xy} \end{pmatrix} = \frac{A^2}{8\pi}(r-r^{-1}) \begin{pmatrix} \sin(2\theta) & -\cos(2\theta) \\ -\cos(2\theta) & -\sin(2\theta) \end{pmatrix}.
\end{equation}

\subsection{Model equations}
\label{sec:elliptic_model_eqs}

The motion of well-separated elliptical vortex patches is approximated by the Hamiltonian elliptic model of Melander, Zabusky \& Styczek \cite{MZS}. This model approximates the evolution of a collection of $N$ vortex patches with vorticity $\omega_i$, area $A_i$, centroid $\mathbf{x}_i$  and elliptic shape $(r_i,\theta_i)$. The associated Hamiltonian is given by the excess energy of the system \cite{vortex_dynamics}
\begin{equation}
 H = \sum_{i=1}^N \underbrace{- \frac{\omega_i}{2} \int_{A_i} \psi(\hat{\mathbf{x}})\,d^2\hat{\mathbf{x}}}_{\equiv H_i},
\end{equation}
where $\psi(\mathbf{x})$ is a stream function for the flow. Each integral over a vortex patch can be evaluated as
\begin{equation}
\label{eqn:Hamiltonian}
 H_i = \underbrace{\frac{(\omega_i A_i)^2}{8 \pi} \ln\left(\frac{(1+r_i)^2}{4r_i}\right)}_{\mbox{self}} + \underbrace{\sum_{j \neq i}^N \frac{(\omega_i A_i)(\omega_j A_j)}{4 \pi}\ln(R_{ij})}_{\mbox{point}} -\underbrace{\sum_{j \neq i}^N \frac{\omega_i}{2} \mathbf{p}_i\mathbf{:}\mathbf{E}_{ij}}_{\mbox{elliptic}},
\end{equation}
where
\begin{equation}
 \mathbf{E}_{ij} = \frac{\omega_j A_j}{2 \pi R_{ij}^2} \begin{pmatrix} \sin(2 \theta_{ij}) & -\cos(2 \theta_{ij}) \\ -\cos(2 \theta_{ij})  & -\sin(2 \theta_{ij}) \end{pmatrix}
\end{equation}
is the rate of strain at the centre of patch $i$ due to the leading-order point vortex contribution from patch $j$, $R_{ij} \equiv |\mathbf{x}_i-\mathbf{x}_j|$ is the distance between the centres of patches $i$ and $j$, and $\theta_{ij}$ is the direction of $\mathbf{x}_i-\mathbf{x}_j$. Terms of order $R_{ij}^{-3}$ and higher are neglected in the calculation of the Hamiltonian. This approximation is valid provided $R_{ij}$ is large compared to the longest dimension of patches $i$ and $j$.

The Hamiltonian consists of three parts: a `self' energy, a `point' energy and an `elliptic' energy. The self energy is associated with the self-induced rotation of the vortex patches. The point energy is associated with the leading-order motion of the patch centroids due to point vortex interactions. The elliptic energy is associated with changes in shape due to strain from other patches and also with the next-order corrections to the motion of the patch centroids. We choose to write the `elliptic' contribution to the Hamiltonian in a slightly different form to that given by Melander, Zabusky \& Styczek. Our form has the advantage of emphasising the role of strain in changing the shape of the ellipse.

The model is completed by the evolution equations
\begin{subequations}
\label{eqn:evolution}
\begin{align}
 \dot{y}_i &= \frac{1}{\omega_i A_i}\frac{\partial H}{\partial x_i} \\
 \dot{x}_i &= -\frac{1}{\omega_i A_i}\frac{\partial H}{\partial y_i} \\
 \dot{\theta}_i &= -\frac{8 \pi}{\omega_i A_i^2}\frac{r_i^2}{1-r_i^2} \frac{\partial H}{\partial r_i} \\
 \dot{r}_i &= \frac{8 \pi}{\omega_i A_i^2}\frac{r_i^2}{1-r_i^2} \frac{\partial H}{\partial \theta_i}.
\end{align}
\end{subequations}
Melander, Zabusky \& Styczek derive these equations by inspection from their approximation of the Euler equations. Alternatively, they can be derived directly from the Hamiltonian formulation of the Euler equations (see Meacham, Morrison \& Flierl \cite{Hamiltonian_reduction}).

We note that the model can be easily extended to include the effect of an irrotational background flow. The additional excess energy due to an irrotational background flow with stream function $\Psi$ is given by
\begin{equation}
\label{eqn:background_H}
 H_{b} = - \sum_{i=1}^N \omega_i \int_{A_i} \Psi(\hat{\mathbf{x}})\,d^2\hat{\mathbf{x}}.
\end{equation}
The elliptic approximation is then obtained by evaluating these integrals and neglecting any moments that are higher than second order.

A vital assumption in the use of this Hamiltonian model is that the effects of viscosity can be neglected. Viscosity causes the patch vorticity to diffuse outwards and leads to the formation of viscous boundary layers along fluid boundaries; but perhaps most critically, it breaks the Hamiltonian nature of the system. Such effects can be neglected, and the system treated as Hamiltonian, provided the appropriate Reynolds numbers are large. A patch of vorticity $\omega$ and typical dimension $R$ generates a typical velocity $\omega R$. Thus diffusion of the patch viscosity can be neglected provided the Reynolds number $\omega R^2/\nu$ is large. The effects of viscosity at fluid boundaries can also be neglected provided another Reynolds number $\omega R S/\nu$, based on the separation $S$ between the patch and the boundary, is also large.

\subsection{Contour dynamics}

The elliptic model is only valid provided the vortex patches, or equivalently, the patch and boundary, are sufficiently far apart. In reality, this separation does not have to be very large for the model to be a very good approximation.  Nevertheless, it is informative to compare the model against the full solution. This full solution can be calculated relatively efficiently via the method of contour dynamics \cite{CD} as briefly described below.

The stream function due to a patch of uniform vorticity is given by an integral, over the area of the patch, of the point vortex Green's function
\begin{equation}
 \psi(\mathbf{x}) = -\frac{\omega}{2 \pi} \int_A \ln(|\mathbf{x}-\hat{\mathbf{x}}|)\,d^2\hat{\mathbf{x}}.
\end{equation}
To calculate the fluid velocity due to this patch, derivatives of the stream function must be evaluated. In the contour dynamics method, the divergence theorem is used to convert the expression for these derivatives from an integral over the area of the patch to an integral around the boundary of the patch:
\begin{align}
 \nabla \psi(\mathbf{x}) &= \frac{\omega}{2 \pi} \int_A \nabla_{\hat{\mathbf{x}}} \ln(|\mathbf{x}-\hat{\mathbf{x}}|)\,d^2\hat{\mathbf{x}} \nonumber \\
&= \frac{\omega}{2 \pi} \int_{\partial A} \ln(|\mathbf{x}-\hat{\mathbf{x}}|) \,d\hat{\mathbf{n}}.
\end{align}
Thus the velocity of points on the vortex patch boundaries, which are required to evolve the patch positions, can be calculated by integrating around only the boundaries of the vortex patches.

\section{Straight boundary}
\label{sec:straight}

\begin{figure}
 \centering
 \includegraphics[width=0.7\textwidth]{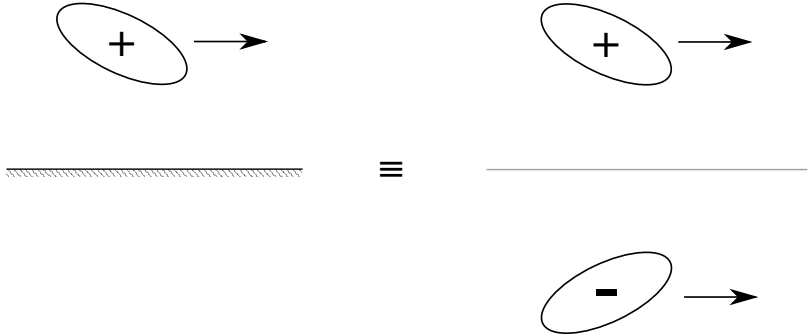}
 \caption{The motion of a vortex patch above a straight boundary is equivalent, via the method of images, to that of a vortex patch and an image vortex patch, of opposite vorticity, obtained by a reflection in the boundary.}
 \label{fig:wall}
\end{figure}

The simplest interaction of a vortex patch with a boundary is the interaction with a straight boundary. The motion of a vortex patch in the half-plane $y \geq 0$ above a boundary at $y = 0$ is equivalent, via the method of images, to the motion of a vortex patch lying in $y \geq 0$ combined with its mirror image in $y=0$. This image lies in the half-plane $y \leq 0$ and has the opposite sense of vorticity (Figure~\ref{fig:wall}). The Hamiltonian for this system can be constructed using Equation (\ref{eqn:Hamiltonian}). The resulting Hamiltonian is eight-dimensional with four parameters describing each of the original vortex patch and its image. The Hamiltonian governing only the motion of the original patch can be obtained by writing the parameters of the image patch in terms of those of the original patch, and by considering only the excess energy associated with the half-plane $y \geq 0$, which is simply half the excess energy of the full plane. Thus the Hamiltonian for the original patch is
\begin{equation}
\label{eqn:coast}
H(y_c,r,\theta) = \frac{(\omega A)^2}{8 \pi} \ln\left(\frac{(1+r)^2}{4r}\right) - \frac{(\omega A)^2}{4 \pi} \ln( 2y_c ) - \frac{\omega}{2} \mathbf{p}(r,\theta) \mathbf{ : E}(y_c),
\end{equation}
where
\begin{equation}
 \mathbf{E}(y) = -\frac{A \omega}{8 \pi y^2} \begin{pmatrix} 0 & 1 \\ 1 & 0 \end{pmatrix}.
\end{equation}
Due to the translational symmetry of the geometry, the Hamiltonian is independent of $x_c$. This symmetry implies the existence of an additional conserved quantity, which is the distance $y_c$ from the boundary of the centroid of the ellipse. For the four dimensional elliptic model, the presence of an additional conserved quantity implies that the motion is integrable. Furthermore, since $y_c$ is constant, the evolution of the elliptic shape is simply that of an elliptic vortex patch in a constant straining background flow. The steady solutions for an elliptic vortex patch in constant strain were found by Moore \& Saffman \cite{steady_ellipse}. The unsteady case, which is of interest here, was later analysed by Kida\cite{Kida} (who actually analysed the more general problem of an ellipse in shear flow). We can recover his result for the subcase where there is no background vorticity by substituting in for $\mathbf{p}$ and $\mathbf{E}$ in Equation~(\ref{eqn:coast}) to obtain
\begin{equation}
 \cos(2 \theta) = \frac{\omega}{e} \frac{r}{r^2-1}\ln\left(\frac{(r+1)^2}{4sr}\right),
\end{equation}
where $e \equiv (A \omega)/(8 \pi y_c^2)$ is the strength of the straining flow, and $s$ is some function of the constants $H$ and $y_c$.

\begin{figure}
 \centering
 \includegraphics[width=0.7\textwidth]{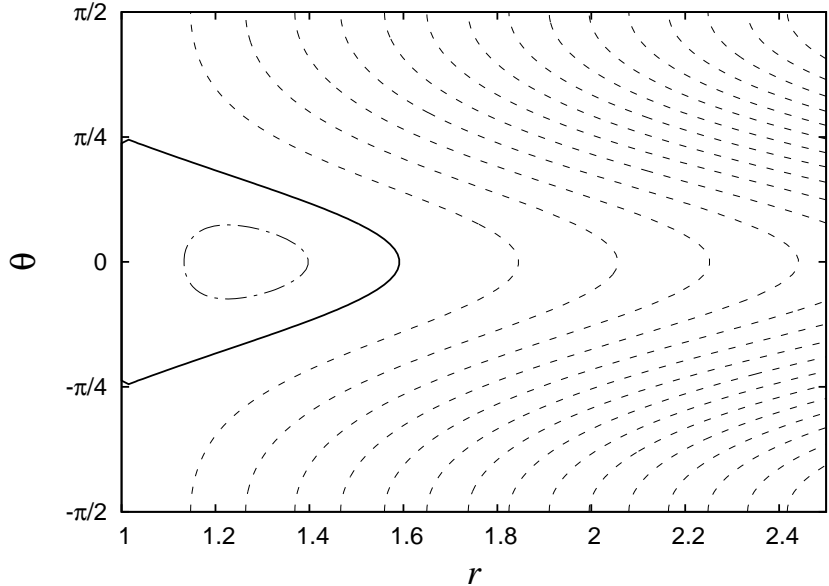}
 \caption{Possible motions in the $r$-$\theta$ plane as predicted by the elliptic model with $e/\omega = 1/18$, corresponding to $y_c/R = 1.5$. The contour that intersects $r = 1$, at which point $\theta$ is undefined, represents the motion of an initially circular patch (solid line).  Any contour within this circular motion contour represents a nutating motion (dot-dashed line). Contours outside the circular motion contour represent vortex patches undergoing rotation (dashed lines). For sufficiently large values of $r$, larger than those shown here, vortex patches are extended indefinitely.}
 \label{fig:Kida}
\end{figure}

Kida found that there are three fundamental types of motion possible: indefinite extension, in which the long axis of the ellipse aligns itself with the axis of extension of the strain ($\theta = -\pi/4$) and $r \rightarrow \infty$; rotation, in which $r$ is bounded and $\theta$ varies monotonically; and nutation, in which $r$ is bounded and $\theta$ oscillates about $\theta = 0$. For sufficiently strong strain, $e/\omega > 0.150$, the only possibility is indefinite extension. For $0.123 < e/\omega < 0.150$ both extension and nutation are possible. Then for sufficiently weak strain, $e/\omega < 0.123$, all three types of motion are possible.  Figure~\ref{fig:Kida} shows possible motions in the $r$-$\theta$ plane when $e/\omega = 1/18 = 0.056$. The closed contour represents a nutating motion. The contours which are monotonic in $\theta$ represent rotating motions. In between these two types of motion lies the motion of an initially circular patch. These three possibilities were shown in Meacham et al.\ \cite{Hamiltonian_reduction} to be conveniently determined by the intersection of a Casimir hyperboloid with the energy surface, just as the motion of the free rigid body is determined by the intersection of the angular momentum sphere with the energy ellipsoid.

\begin{figure}
 \centering
 \includegraphics[width=0.7\textwidth]{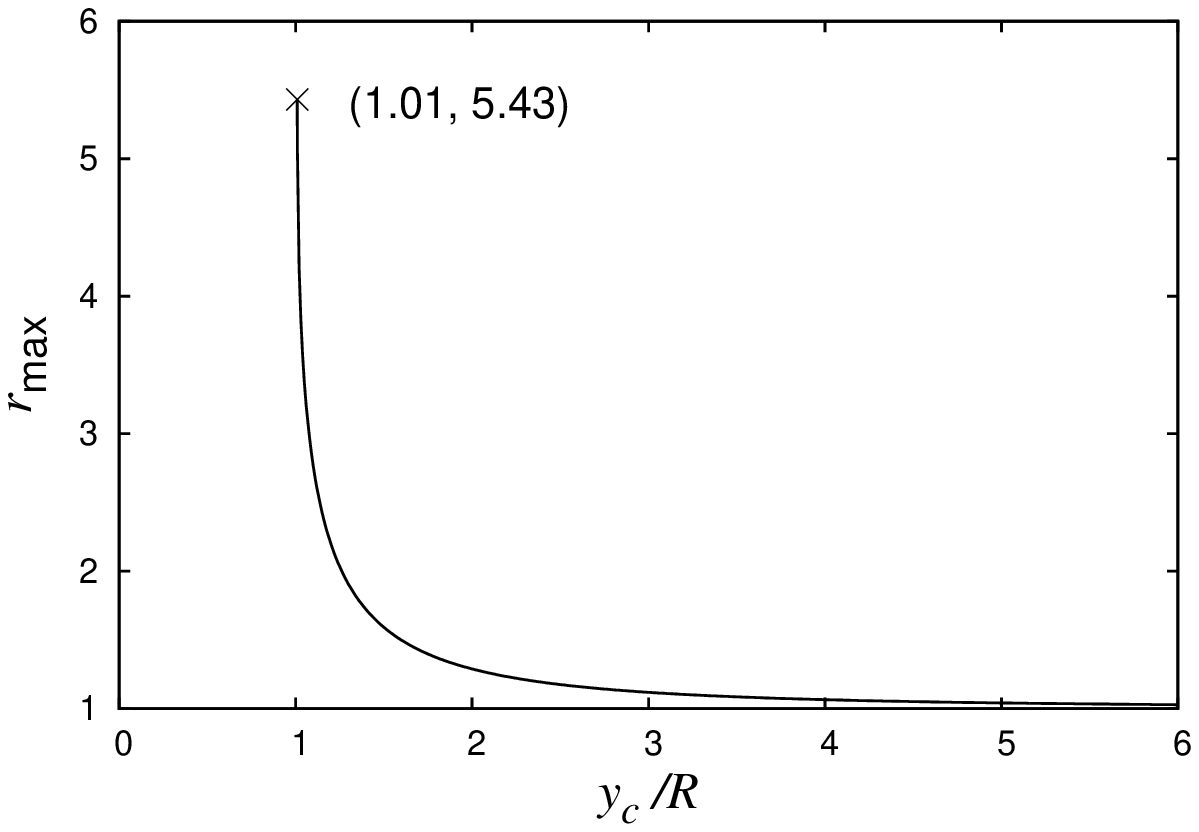}
 \caption{Maximum deformation $r_{\rm max}$ of an initially circular patch of vorticity in terms of the distance of its centroid from the boundary, as given by the elliptic model. The model predicts that the patch is extended indefinitely when $y_c/R < 1.01$.}
 \label{fig:wall_distortion}
\end{figure}

A quantitative measure of how much a boundary deforms vortex patches is the maximum deformation, as measured by $r$, experienced by an intially circular vortex patch.  An initially circular vortex patch, of radius $R$, will be indefinitely extended if $y_c/R < 1.01$ (corresponding to  $e/\omega > 0.123$) or return to its original circular shape in a periodic motion if $y_c/R > 1.01$. For the latter case, Figure~\ref{fig:wall_distortion} shows the maximum deformation $r_{\rm max}$ of the intially circular vortex patch in terms of the distance of its centroid from the boundary. The maximum deformation falls off rapidly with distance from the boundary.  This fall-off is a consequence of the $1/y_c^2$ decay of the straining flow acting on the patch as $y_c$ increases.

\subsection{Model validation}

\begin{figure}
  \centering
  \includegraphics[width=\textwidth]{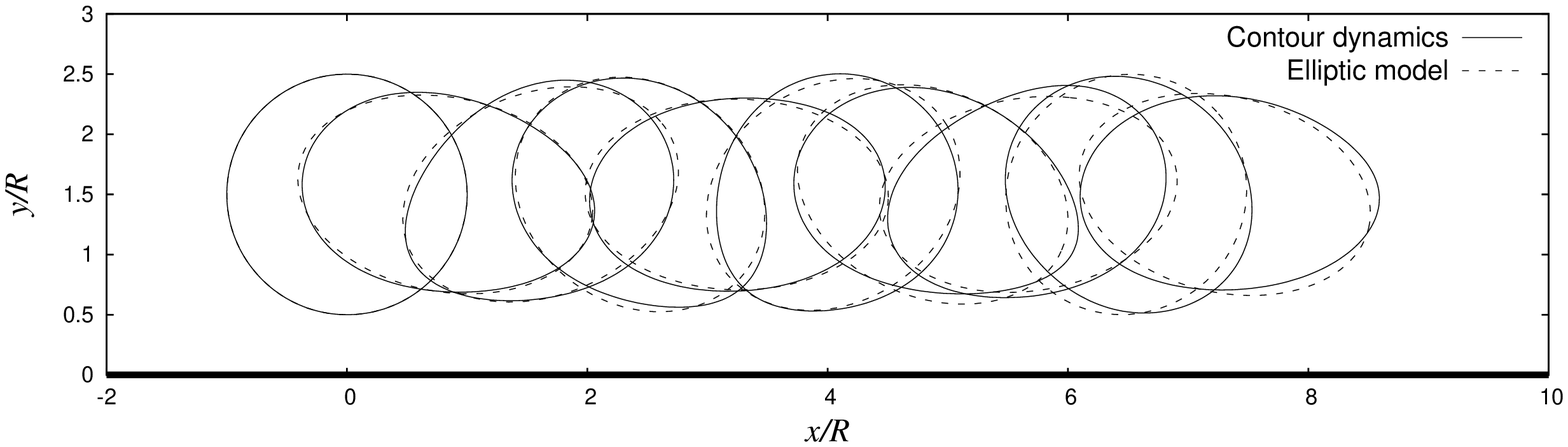}
  \caption{Motion of an initially circular patch with centroid at $y_c = 1.5 R$.  Patch motion is from left to right at time intervals of $5/\omega$. Contour dynamics solution is given by solid lines, and the elliptic approximation by dashed lines.}
  \label{fig:near_wall}
\end{figure}

\begin{figure}
 \centering
 \includegraphics[width=\textwidth]{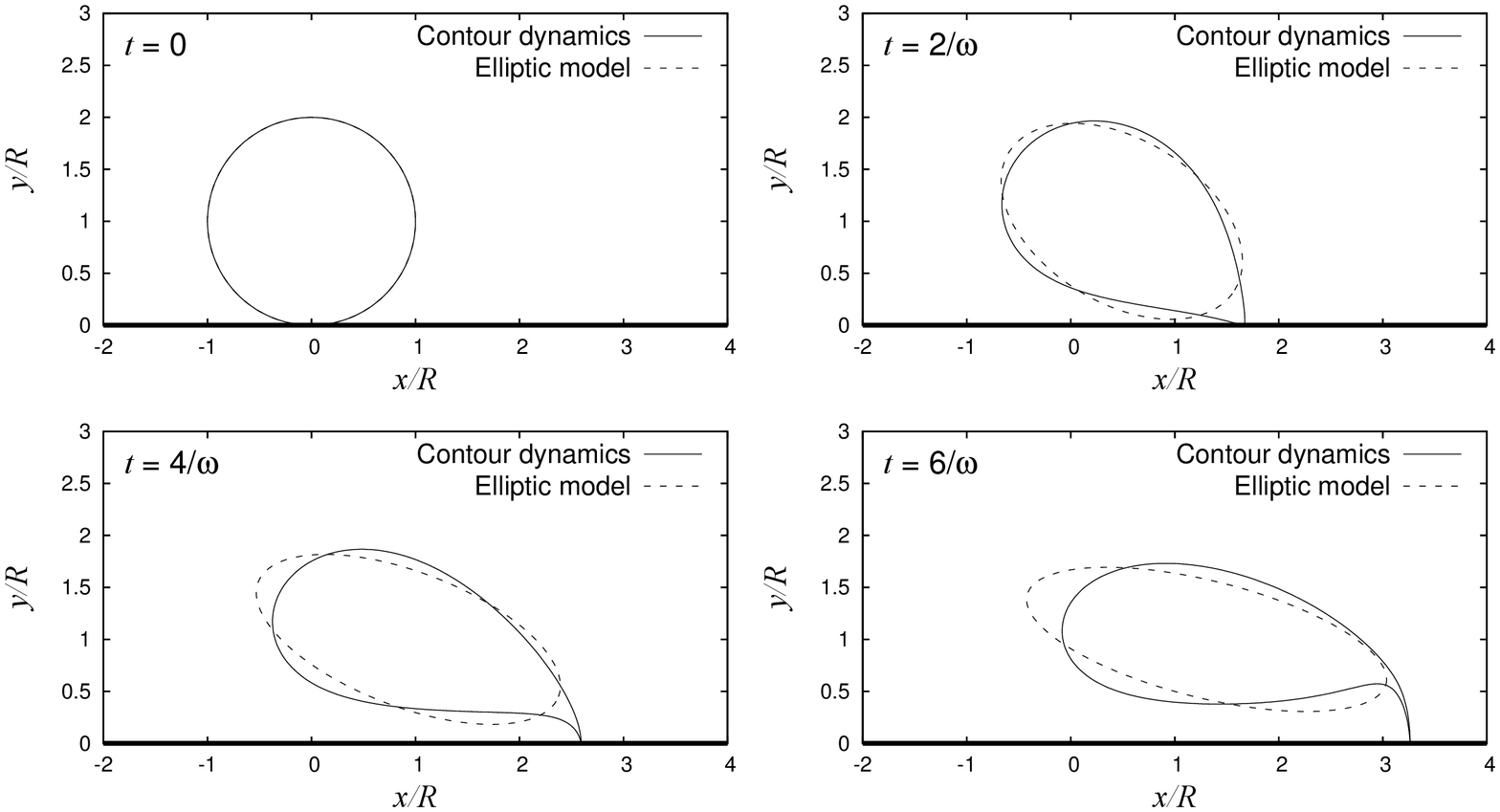}
 \caption{Even when the elliptic model is not valid, it still gives a good approximation to the region occupied by the vortex patch at short times. Here we show the contour dynamics and elliptic model solutions with $y_c = R$ at time intervals of $2/\omega$.}
 \label{fig:wall_y1}
\end{figure}

The elliptic model is only valid if the vortex patch and its image are well separated, but in reality, it provides a good approximation to the motion of an initially circular vortex patch even when it is remarkably close to the boundary. Figure~\ref{fig:near_wall} shows the evolution of such a vortex patch with $y_c = 1.5 R$ both under the elliptic model and the full contour dynamics solution. There is very good agreement between the two, with the most notable difference being a slight phase drift with time.

Even for $y_c = R$, when the contour dynamics solution shows that the vortex patch is stretched out into a non-elliptical shape, the elliptic model gives a remarkably good approximation to the overall region occupied by the vortex patch at short times (Figure~\ref{fig:wall_y1}).

\section{Motion around a corner}
\label{sec:corner}

\begin{figure}
 \centering
 \includegraphics[width=0.6\textwidth]{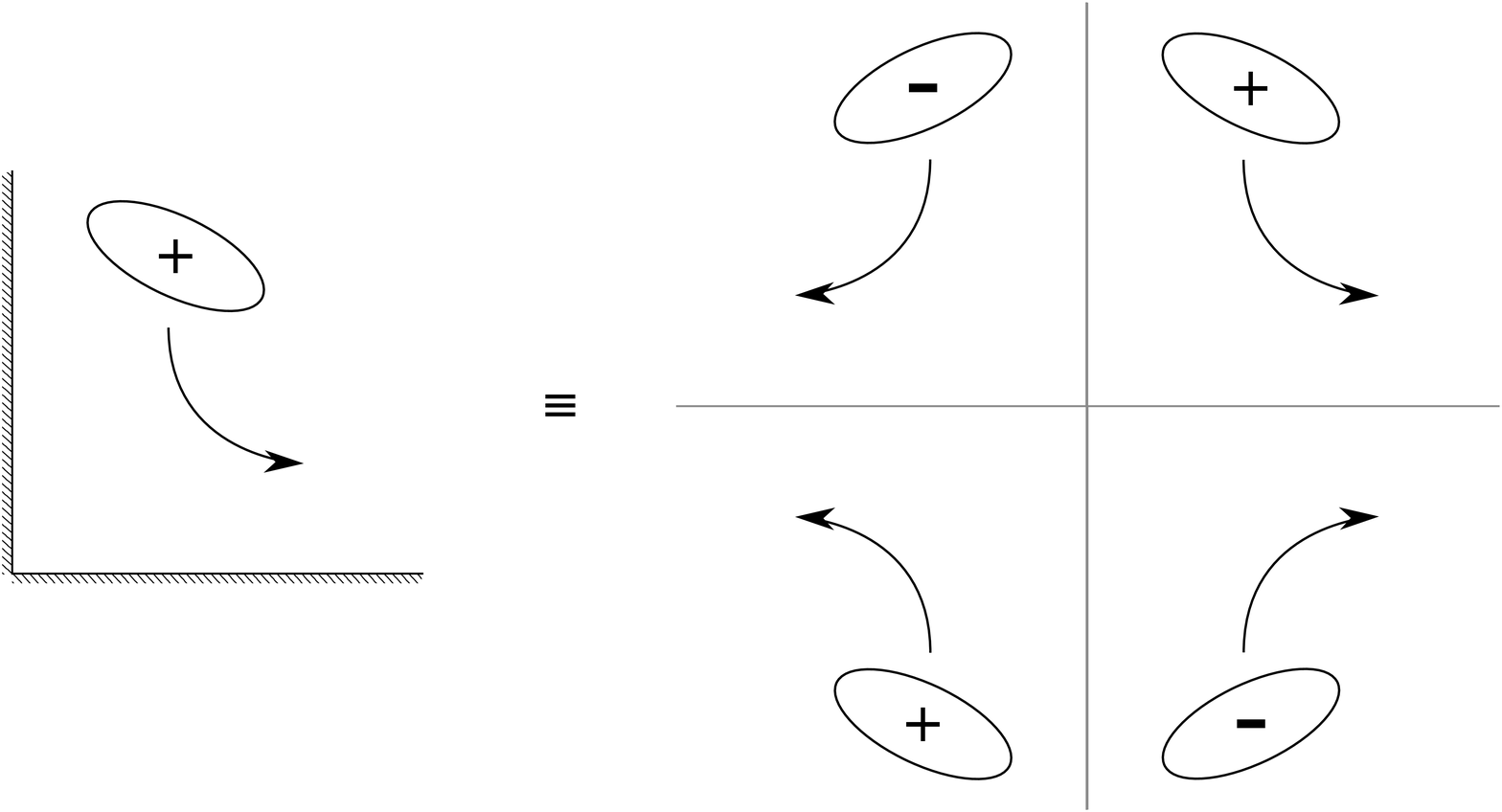}
 \caption{The motion of a vortex patch around a corner is equivalent, via the method of images, to that of a vortex patch and its image vortex patches obtained by reflection in the boundaries $x = 0$ and $y = 0$. The two image vortex patches obtained from a single reflection have the opposite vorticity to the original patch, the third image patch has the same vorticity.}
 \label{fig:corner_images}
\end{figure}

Since the motion above a straight boundary is periodic, the only way that large deformations to the vortex patch can occur is if the separation of the patch from the boundary is comparable to its radius. For more complicated boundaries, we might expect deformations to the patch shape to accumulate over time. In which case, even vortex patches whose separation from the boundary is large compared to their radius could eventually accumulate large deformations. This motivates us to consider the motion of a vortex patch in a quarter plane ($x \geq 0$ and $y \geq 0$).

Since a quarter plane does not possess the translational symmetry of the half plane, the motion of a vortex patch within a quarter plane is unlikely to be integrable under the elliptic model. This could be determined by a Melnikov analysis, although such analyses in the present context can have special difficulties. (See Ngan, Meacham \& Morrison\cite{ngan} and references therein.) Nevertheless, the geometry is still simple enough to allow the use of the method of images. In this case, two reflections must be made: one in $y = 0$, and one in $x = 0$ (Figure~\ref{fig:corner_images}). This gives three image vortex patches, two of which have opposite vorticity to the original patch. The Hamiltonian for the motion of the original patch (a quarter of the excess energy for the full plane with images) is given by
\begin{equation}
 H = \frac{(\omega A)^2}{8 \pi} \ln \left( \frac{(1+r)^2}{4r} \right) - \frac{(\omega A)^2}{4 \pi} \ln \left( \frac{2x_cy_c}{\sqrt{x_c^2+y_c^2}} \right) - \frac{\omega}{2} \mathbf{p}(r,\theta) \mathbf{:E}(x_c,y_c),
\end{equation}
where the strain rate
\begin{equation}
 \mathbf{E}(x,y) = \frac{A \omega}{8 \pi} \begin{pmatrix} \frac{2xy}{(x^2+y^2)^2}  & \frac{1}{x^2}-\frac{1}{y^2}+\frac{y^2-x^2}{(x^2+y^2)^2}& \\  \frac{1}{x^2}-\frac{1}{y^2}+\frac{y^2-x^2}{(x^2+y^2)^2}&  -\frac{2xy}{(x^2+y^2)^2} \end{pmatrix}
\end{equation}
now depends on both $x_c$ and $y_c$.

A patch of positive vorticity will propagate from the region $y \gg x$, around the corner, and then towards the region $x \gg y$; and vice-versa for a patch of negative vorticity.  When either $x_c \gg y_c$ or $y_c \gg x_c$ the evolution of the patch is well approximated by that of a patch above a straight boundary, for which the distance of the centroid from the boundary is fixed.  We define $y_{\infty}$ and $x_{\infty}$ to be the limits of $y_c$ and $x_c$ as $x_c \rightarrow \infty$ and $y_c \rightarrow \infty$ respectively.

\begin{figure}
 \centering
 \includegraphics[width=\textwidth]{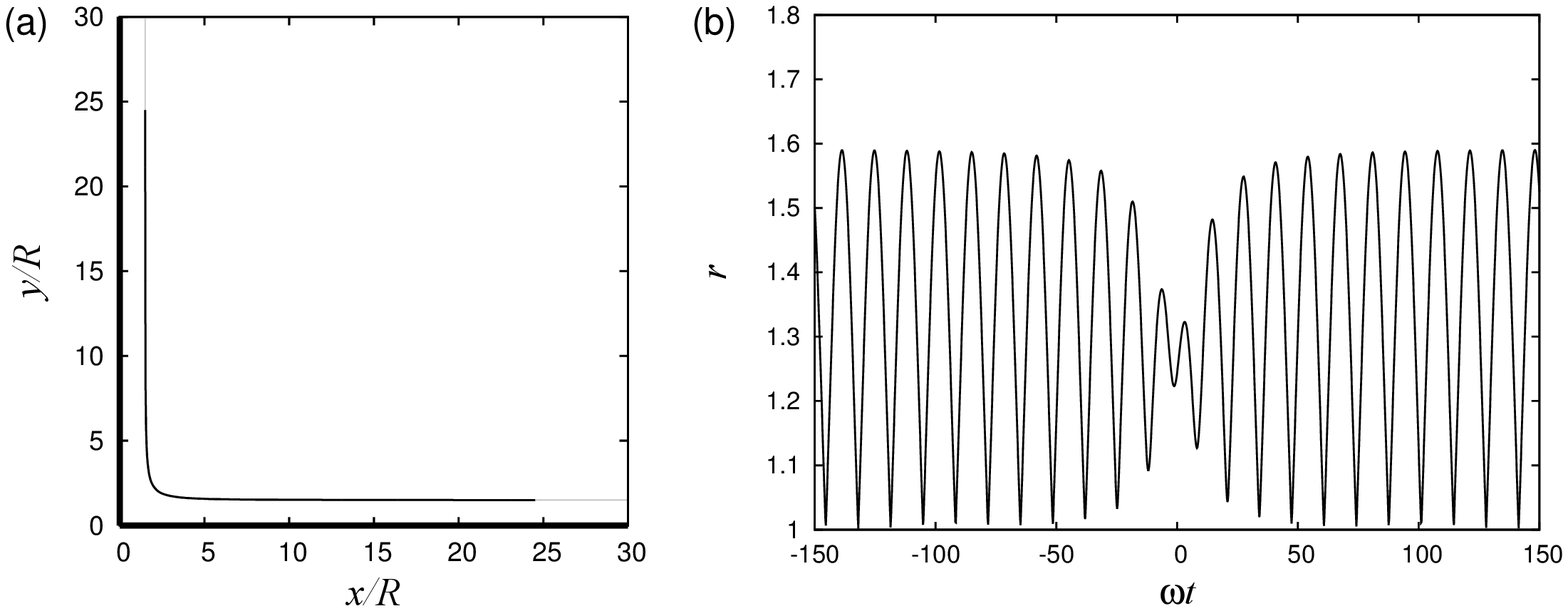}
 \caption{Evolution of an intially circular vortex patch in a quarter plane with initial centroid $(1.5R,100R)$. The time origin $t = 0$ is defined to occur when $x_c = y_c$. (a) Path of the patch centroid. (b) Variation of elliptic shape with time. Far away from the corner, the patch undergoes a periodic motion associated with the evolution of a patch above a straight boundary. The periodic evolution of the patch shape before and after the corner is almost exactly the same.}
\label{fig:corner}
\end{figure}

Figure~\ref{fig:corner} shows the evolution of an initially circular vortex patch with $x_{\infty} = 1.5R$ moving around the corner. The most notable feature of this evolution is the similarity of the periodic variation of $r$ with $t$ before and after the interaction with the corner region.  The motion afterwards is almost exactly the same, periodically circular, motion that the vortex patch was executing before it approached the corner. This similarity suggests that $x_{\infty} \approx y_{\infty}$, and indeed we find that $|x_{\infty}-y_{\infty}| = 4\times 10^{-6}R$. The approximate equality between $x_{\infty}$ and $y_{\infty}$, and the corresponding similar periodic evolution of $r$, are generic. It appears to occur for all initially circular patches with $x_\infty \geq 1.01R$, i.e.\ all initially circular patches that are not indefinitely extended when at a similar separation from a straight boundary (Figure~\ref{fig:wall_distortion}). The explanation for this phenomenon lies in the existence of an adiabatic invariant.

\subsection{Adiabatic invariance}

Adiabatic invariance occurs for two-dimensional Hamiltonian systems within a slowly varying environment. Due to the separation of time scales between the motion of the system and the variation of its environment there exists an (approximate) invariant of the motion referred to as the adiabatic invariant\cite{adiabatic}. Strictly speaking, the adiabatic invariant is not an invariant but rather a quantity which varies by only a very small amount. It is constant to all orders in $\epsilon$, where $\epsilon$ is defined to be the small ratio of the time scale of the system to the time scale on which the environment varies, but can vary by an exponentially small amount.

Returning to an elliptic vortex patch in a quarter plane, we observe that there are two time scales associated with the motion of the vortex around the corner: the time scale on which the patch rotates, $T_{\rm rot}$; and the time time scale on which the strain experienced by the patch varies, $T_{\rm strain}$. From Figure~\ref{fig:corner}(b), we see that $T_{\rm rot}$ is the shorter of these two time scales. The variation in strain is largely confined to the period $-40 \leq \omega t \leq 40$ but during this period the patch rotates multiple times. This separation of time scales, while not an order of magnitude, is still sufficient to lead to the phenomenon of adiabatic invariance. (That an elliptic vortex patch in a slowly varying strain can behave adiabatically was first observed by Legras~\&~Dritschel\cite{vortex_stripping_adiabatic}.)

\begin{figure}
 \centering
 \includegraphics[width=\textwidth]{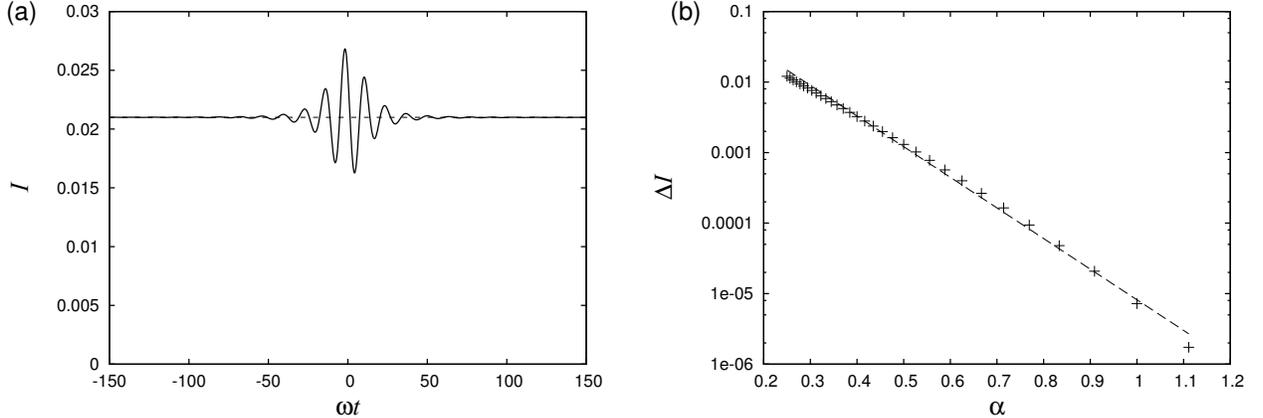}
 \caption{(a) Evolution of the action $I$ for the motion shown in Figure \ref{fig:corner}. The values of $I$ before and after the corner interaction are almost exactly the same. (b) The difference in action $\Delta I$ before and after the corner is shown for patches whose centroid speed is modified by a factor of $\alpha^{-1}$. These values of $\Delta I$ are calculated from a root mean square average over $q$ of the difference in action for all vortex patches with the same initial action and energy (and consequently separation from the wall) as the patch shown in Figure \ref{fig:corner}. The difference $\Delta I$ decreases exponentially as $\alpha$ is increased (corresponding to $\epsilon$ being decreased), which is consistent with adiabatic behaviour.}
 \label{fig:corner_action}
\end{figure}

The separation of time scales allows us to consider the evolution of the vortex patch shape as a two-dimensional dynamical system with a slowly varying parameter instead of as part of the full four-dimensional elliptic model. This slowly varying parameter is the strain rate $\mathbf{E}$ experienced by the vortex patch. The Hamiltonian for the two-dimensional system is given by
\begin{equation}
 H_{\rm shape}(r,\theta,t) \equiv \frac{(\omega A)^2}{8 \pi} \ln \left( \frac{(1+r)^2}{4r} \right) - \frac{\omega}{2} \mathbf{p}(r,\theta) \mathbf{:E}( \epsilon \omega t).
\end{equation}
We write $\mathbf{E} = \mathbf{E}( \epsilon \omega t)$, where $\epsilon \ll 1$, to emphasise that the variation in strain rate is slow compared to the rotation of the patch. The theory of adiabatic invariance applies to this system and implies the existence of an adiabatic invariant that is constant to all orders in $\epsilon$. Furthermore, when $\mathbf{E}$ is constant, the value of this adiabatic invariant is simply given by the action
\begin{equation}
 I(p,q,\mathbf{E}) \equiv \oint p \,dq,
\end{equation}
where
\begin{equation}
 p \equiv \frac{\omega A^2}{16 \pi}\frac{(r-1)^2}{r} \quad \mbox{and} \quad q \equiv 2 \theta
\end{equation}
are canonical variables for the Hamiltonian system \cite{MZS}.

Sufficiently far before and after the corner, the strain experienced by the patch, and hence the action, are constant. We define $I_-$ to be the action before, and $I_+$ to be the action afterwards. The existence of an adiabatic invariant implies that the difference in action, $\Delta I \equiv I_+ - I_-$, should be constant to all orders in $\epsilon$. Figure~\ref{fig:corner_action}(a) shows the evolution of $I$ for the same motion as shown in Figure~\ref{fig:corner}. During the interaction with the corner region $I$ varies, but as $t \rightarrow \pm \infty$ it tends to constant values, and the difference in these values is small ($\Delta I \ll I$), which is consistent with the existence of an adiabatic invariant.

To confirm that $\Delta I$ has the constant to all orders in $\epsilon$ behaviour associated with the existence of an adiabatic invariant, we must analyse the variation of $\Delta I$ with $\epsilon$. The rate at which $\mathbf{E}$ varies, and hence $\epsilon$, is determined by the speed at which the vortex patch moves through the corner. Therefore, we consider the motion of vortex patches with $\dot{r}$ and $\dot{\theta}$ as before (Equation~\ref{eqn:evolution}), but with $\dot{x}$ and $\dot{y}$ modified by a factor of $\alpha^{-1}$. By varying $\alpha$, we can vary the separation of the time scales $\epsilon$, which will vary as $\alpha^{-1}$.  The resulting variation of $\Delta I$ with $\alpha$ is shown in Figure~\ref{fig:corner_action}(b).  There is an exponential decrease as $\alpha$ is increased, which confirms that $\Delta I$ is constant to all orders in $\epsilon$.

\medskip

The phenomenon of adiabatic invariance will not be limited to this particular boundary shape, but rather can be expected to occur for any boundary shape along which the motion of vortex patches satisfies $T_{\rm strain} \gg T_{\rm rot}$. Adiabatic invariance could be broken by a boundary with variations in shape on a scale much less than the dimensions of the vortex patch, since $T_{\rm strain}$ will be greatly reduced for such a boundary. However, fluctuations in the straining flow due to short scale boundary variations can be expected to decay exponentially with distance from the boundary on the same scale as the boundary variations (cf.\ solutions of Laplace's equation with sinusoidal boundary conditions). Thus the straining flow generated by such boundary variations will be exponentially small for any vortex patch whose separation from the boundary is large compared to its radius.

We conclude that there are two possible scenarios for vortex patches in the vicinity of a boundary (in the absence of a background flow): either the separation of the vortex patch from the boundary is comparable to its radius, in which case the vortex patch will undergo significant (order one) deformations to its shape (Figure~\ref{fig:wall_distortion}); or the separation of the vortex patch from the boundary is large compared to its radius, in which case the initial deformation to the shape of the vortex patch is small, and, due to the existence of an adiabatic invariant, will remain small. In particular, every time the patch returns to a straight section of boundary, its separation from the boundary and the periodically varying evolution of its shape must be very nearly the same.

\section{Chaotic behaviour of the elliptic model}
\label{sec:chaos}

\begin{figure}
 \centering
 \includegraphics[width=\textwidth]{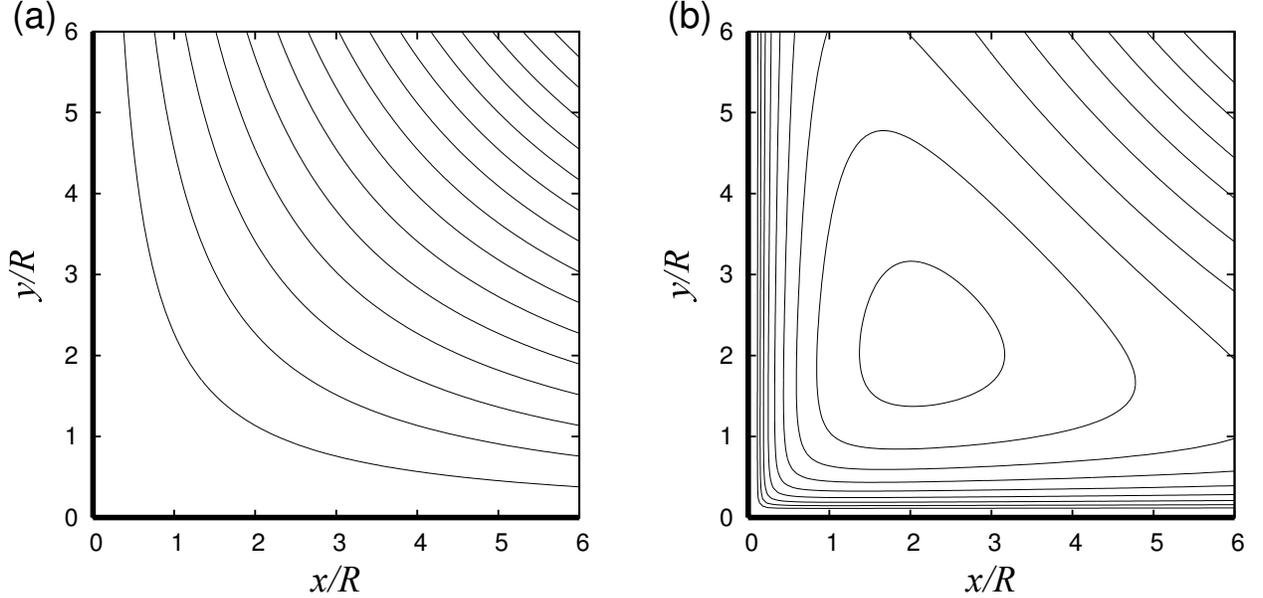}
 \caption{(a) Streamlines for the irrotational background flow $\Psi(x,y) = \beta x y$. (b) Paths of patch centroids under the point vortex approximation for the case $\beta = -0.028 \omega$. Whenever $\beta/\omega < 0$, a stagnation point exists, and vortex patches are trapped in the corner region.}
\label{fig:corner_strain}
\end{figure}

The elliptic model for the motion of a vortex patch in the vicinity of a boundary is four-dimensional, and a generic feature of such four-dimensional dynamical systems is chaotic behaviour. In the examples that we have considered so far, any chaotic behaviour has been suppressed either by the motion being integrable, or by the presence of an adiabatic invariant. To demonstrate that chaotic behaviour is possible under the elliptic model, we briefly consider the motion of a vortex patch trapped in a corner by a background straining flow.

If a pure straining flow, with stream function $\Psi = \beta x y$, is added to the quarter plane geometry of the previous section, then the straining flow can trap vortex patches in the corner provided the sign of $\beta$ is appropriately chosen. Figure~\ref{fig:corner_strain}(a) shows the streamlines for this background flow, and Figure~\ref{fig:corner_strain}(b) shows representative paths of vortex patches approximated as point vortices, which are trapped in the corner. This was previously noted by Llewellyn Smith\cite{lsmith}. The additional excess energy due to the straining flow is 
\begin{equation}
 H_{b} = -\beta \omega A x_c y_c - \frac{\omega}{2} \mathbf{p}(r,\theta) \mathbf{: E}_{b},
\end{equation}
where
\begin{equation}
 \mathbf{E}_{b} = \beta \begin{pmatrix} 1  & 0 \\  0&  -1\end{pmatrix}
\end{equation}
is the strain rate of the background flow.

\begin{figure}
 \centering
 \includegraphics[width=\textwidth]{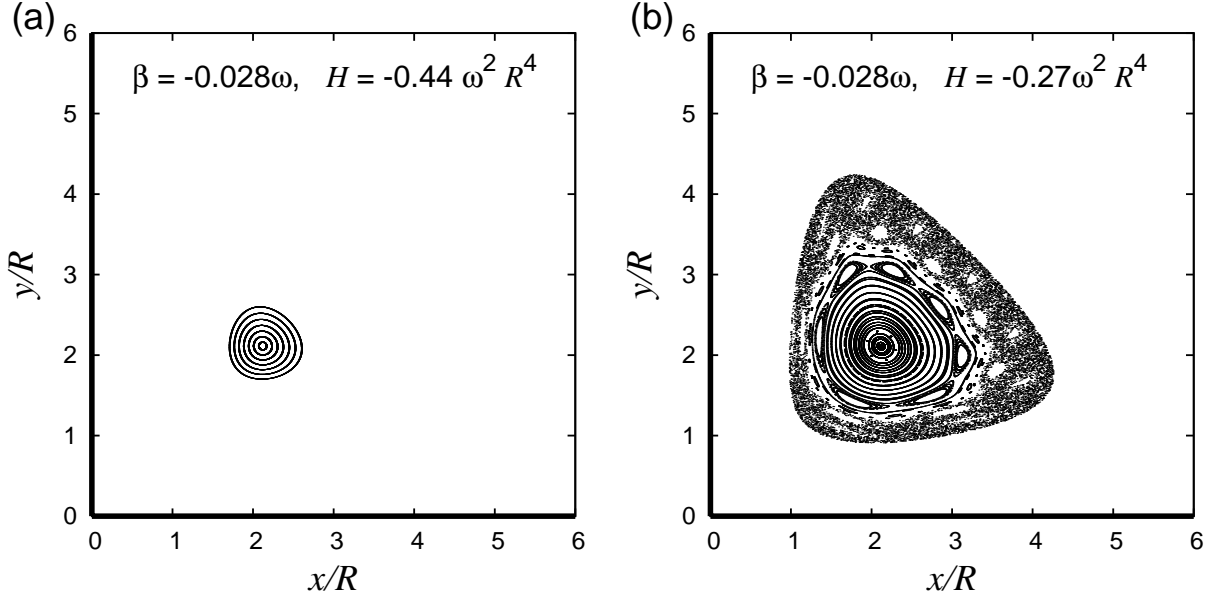}
 \caption{Poincar\'e sections in the $x_c$-$y_c$ plane of forward intersections with $\theta = 0$. The strain rate is $\beta = -0.028 \omega$ and two different values of $H$ are considered. (a) $H/(\omega^2R^4) = -0.44$. The Poincar\'e section closely matches the earlier point vortex paths. (b) $H/(\omega^2R^4) = - 0.27$. The Poincar\'e section reveals chaotic behaviour. Resonances have opened up in some of the previously circular contours, and a `sea of chaos' has formed around the edge of the region.}
\label{fig:corner_psections}
\end{figure}

The resulting motion of the elliptic vortex patches can be visualised via a Poincar\'e section. Two such Poincar\'e sections are shown in Figure~\ref{fig:corner_psections} for the case $\beta = -0.028\omega$. Each Poincar\'e section visualises the motion of several vortex patches, each with the same excess energy $H$, and depicts the points at which the angle $\theta$ of a vortex patch passed forwards through $\theta = 0$. There is no chaotic behaviour visible for the lower value of $H$ shown in Figure~\ref{fig:corner_psections}(a). However, for the larger value of $H$ shown in Figure~\ref{fig:corner_psections}(b), two signs of chaotic behaviour are clearly visible: resonances and a `sea of chaos' \cite{chaos}. Those patches whose motions form the the sea of chaos still orbit the central stagnation point, but the path of the patch centroid can differ significantly between consecutive orbits. Since the patch energy is constant, there must be a significant change of patch shape associated with this change of centroid position.

The motion shown in Figure~\ref{fig:corner_psections}(b) is actually unphysical due to the intersection of some of the elliptic vortex patches with the boundary of the domain; and we were unable to find any values of $\beta$ and $H$ that exhibited such clear chaotic behaviour without also being unphysical. Nevertheless, these results are informative since they emphasise the significance of the suppression of chaotic behaviour by adiabatic invariance that was observed in the preceding section.

\section{Interaction with an island}
\label{sec:island}

We now consider the deformation of a vortex patch in the vicinity of a circular island. This analysis has particular relevance to the propagation of Meddies in the Atlantic, which have been observed to interact with, and sometimes be broken up by, underwater seamounts \cite{meddies}.

\subsection{No background flow}

We begin with the motion of a vortex patch in the vicinity of a circular island with no background flow. For the case of a point vortex, motion around an island can still be solved via the method of images. A point vortex of strength $\kappa$ located at $\mathbf{x}$ requires an image of strength $-\kappa$ located at
\begin{equation}
 \mathbf{x}_I = \mathbf{x} \frac{a^2}{|\mathbf{x}|^2},
\end{equation}
where $a$ is the radius of the circular island \cite{Lamb}. A side effect of this image is that it induces a circulation of $-\kappa$ around the island. Such circulation is a constant of the motion, and, since we are ultimately interested in vortex patches advected towards the island from infinity, we require it to be zero. The condition of zero circulation can be enforced via the introduction of a second image, of strength $\kappa$, at the centre of the island. The resulting stream function is
\begin{equation}
\label{eqn:island_stream_fn}
 \psi(\mathbf{x}) = \frac{\kappa}{2 \pi} \ln \left( \frac{|\mathbf{x}-\mathbf{x}_I|}{|\mathbf{x}|} \right) = \frac{\kappa}{2 \pi} \ln \left( 1- \frac{a^2}{x^2+y^2} \right) .
\end{equation}

The image system for an elliptical vortex patch of uniform vorticity consists of a non-elliptical region of non-uniform vorticity.  Fortunately, the elliptic model only requires the strain due to the leading-order point vortex image, and thus the Hamiltonian (excess energy in the region $x^2+y^2 \geq a^2$) is easily calculated as
\begin{equation}
 H = \frac{(\omega A)^2}{8 \pi} \ln \left( \frac{(1+r)^2}{4r} \right) - \frac{(\omega A)^2}{4 \pi} \ln \left( 1- \frac{a^2}{x_c^2+y_c^2} \right) - \frac{\omega}{2} \mathbf{p}(r,\theta) \mathbf{:E}(x_c,y_c),
\end{equation}
where the strain rate, calculated from Equation~(\ref{eqn:island_stream_fn}), is given by
\begin{equation}
 \mathbf{E}(x,y) = \frac{\omega A}{2 \pi} \left( 1 - \left(1-\frac{a^2}{x^2+y^2}\right)^{-2} \right) \begin{pmatrix} \frac{2xy}{(x^2+y^2)^2}  & \frac{y^2-x^2}{(x^2+y^2)^2}& \\  \frac{y^2-x^2}{(x^2+y^2)^2}&  -\frac{2xy}{(x^2+y^2)^2} \end{pmatrix}.
\end{equation}
The elliptic model is valid provided the centres of the vortex patch and its image are well-separated compared to the patch radius, i.e.\ $s_c - a^2/s_c \gg R$ where $s_c \equiv (x_c^2+y_c^2)^{1/2}$. As before, this separation does not have to be particularly large before the elliptic model becomes a good approximation.

\begin{figure}
 \centering
 \includegraphics[width=0.7\textwidth]{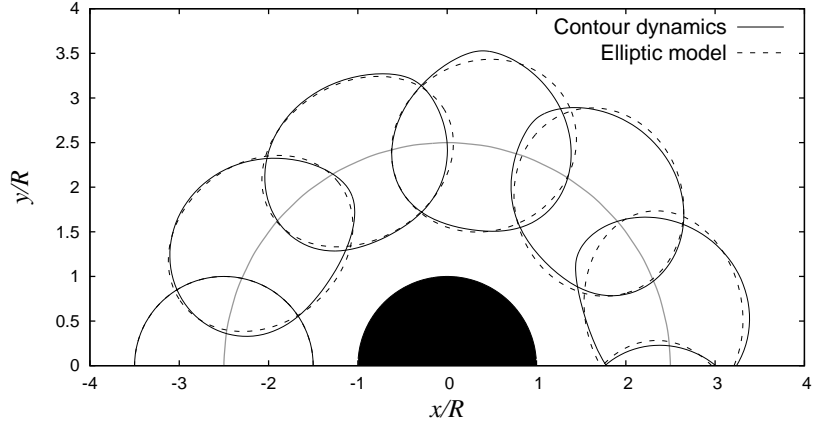}
 \caption{Motion of a vortex patch, initially circular with centroid $(-2.5R,0)$, around an island of radius $a = R$. The motion is visualised by snapshots of the patch location at time intervals of $40/\omega$ for the elliptic model (dashed line) and the full contour dynamics solution (solid line). The path of the centroid under the elliptic model is also shown (grey line).}
 \label{fig:island}
\end{figure}

Since the island geometry has rotational symmetry, there is an additional conserved quantity, which is the angular impulse
\begin{equation}
 \omega \int_A \hat{\mathbf{x}} \mathbf{\cdot} \hat{\mathbf{x}} \,d^2\hat{\mathbf{x}} = \omega \left( A(x_c^2 + y_c^2) + q_{xx}+q_{yy} \right).
\end{equation}
The existence of an additional conserved quantity implies that the motion is integrable under the elliptic model. An example of the evolution of an initially circular vortex patch around an island is shown in Figure~\ref{fig:island}. Also shown is the full contour dynamics solution, which is in good agreement with the elliptic model for $t \leq 80 / \omega$ , but which starts to develop some non-elliptic deformations at later times. (See Appendix~\ref{sec:island_CD} for a derivation of the contour dynamics method for a circular island.)

\begin{figure}
 \centering
 \includegraphics[width=0.7\textwidth]{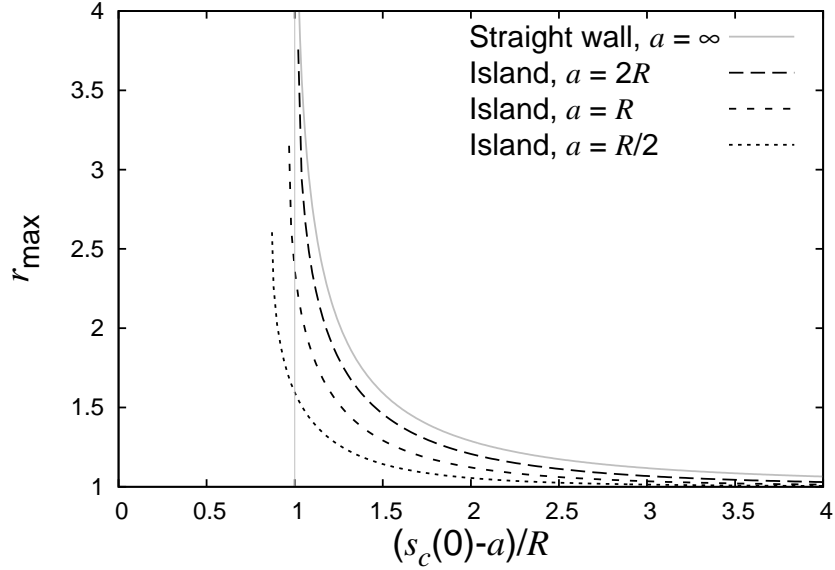}
 \caption{Maximum deformation $r_{\rm max}$ of an initially circular patch of vorticity in terms of the initial separation of its centroid from the boundary, $s_c(0)-a$, as given by the elliptic model. Larger islands lead to more deformation, and, in the limit $a \rightarrow \infty$ the result for deformation above a straight boundary is recovered. Solutions with $s_c(0)-a < R$ are unphysical since the vortex patch overlaps with the island.}
 \label{fig:island_distortion}
\end{figure}

We again use the maximum value of $r$ for an initially circular vortex patch, $r_{\rm max}$, as a measure of the deformation caused by the island.  Figure~\ref{fig:island_distortion} shows this quantity evaluated for island radii of $a = R/2$, $R$ and $2R$ alongside the previous result for a straight boundary ($a = \infty$) for comparison. For a given separation from the boundary, the deformation due to an island is less than that due to a straight boundary, and decreases as the radius of the island is decreased; this is due to a reduction in the magnitude of the straining flow. The strain rate due to an island of radius $a$ has an asymptotic scaling of $a^2/s_c^4$ as $s_c \rightarrow \infty$, which decays faster than the $1/y_c^2$ decay associated with a straight boundary, and which decreases in magnitude as the island radius is decreased. 

\subsection{Background flow}

\begin{figure}
 \centering
 \includegraphics[width=\textwidth]{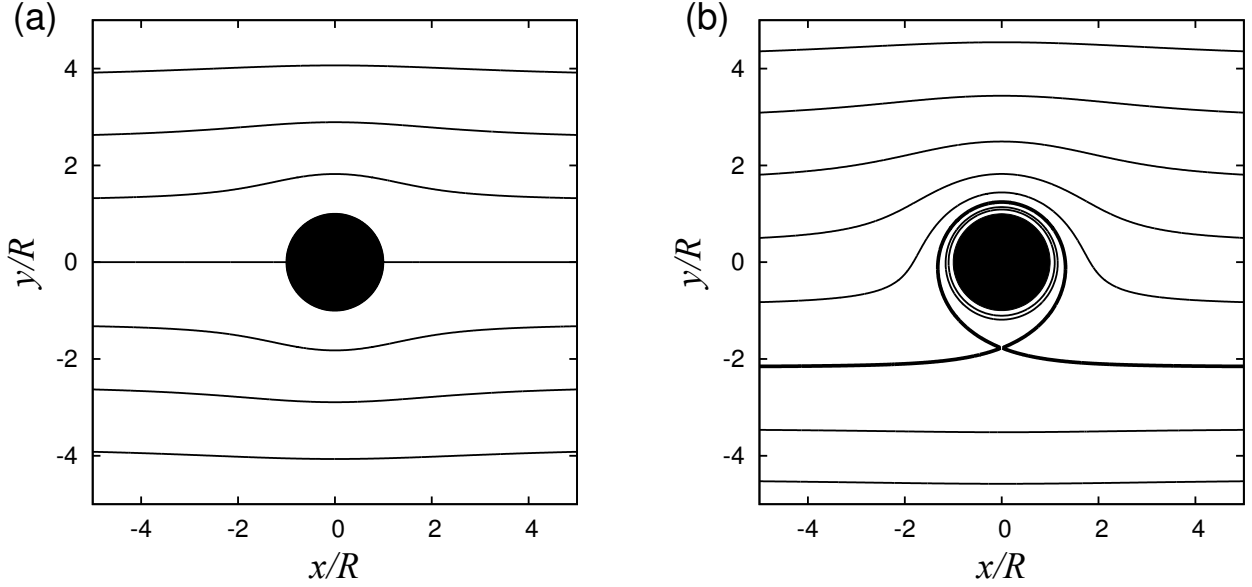}
 \caption{(a) Streamlines for the irrotational background flow $\Psi(x,y) = Uy\left(1-a^2(x^2+y^2)^{-1}\right)$ around an island of size $a = R$. (b) Paths of patch centroid under point vortex approximation with $a = R$ and $U/(\omega R) = 0.1$. The separatrix (heavy line) splits the motion into three regions: vortices passing below the island, vortices passing above the island, and vortices orbitting the island. This separation into three regions is generic and does not depend on the strength of the vortex.}
\label{fig:island_flow}
\end{figure}

The effect of a background flow past the island can be investigated using an irrotational flow with the following stream function:
\begin{equation}
 \Psi(x,y) = Uy\left(1-\frac{a^2}{x^2+y^2}\right).
\end{equation}
This represents a flow which tends to uniform flow away from the island. The associated streamlines are shown in Figure~\ref{fig:island_flow}(a), and a representative example of the resulting point vortex paths is shown in Figure~\ref{fig:island_flow}(b). Point vortices either pass above the island, pass below the island, or are trapped circulating around the island. However, such point vortex paths don't give us much information about the deformation of finite area vortex patches, for which we need the elliptic model.

The additional excess energy due to the background flow (using Equation~(\ref{eqn:background_H}) and neglecting any moments higher than second order) is
\begin{equation}
 H_{b} = -\omega A U y_c \left( 1 - \frac{a^2}{x_c^2+y_c^2} \right) - \frac{\omega}{2} \mathbf{p}(r,\theta) \mathbf{: E}_{b}(x_c,y_c),
\end{equation}
where
\begin{equation}
 \mathbf{E}_{b}(x,y) = \frac{2 U a^2}{(x^2+y^2)^3} \begin{pmatrix} x^3-3y^2x  & 3x^2y-y^3 \\  3x^2y-y^3&  3y^2x-x^3 \end{pmatrix}
\end{equation}
is the strain rate at the patch centroid due to the background flow.

\begin{figure}
\centering
 \includegraphics[width=0.7\textwidth]{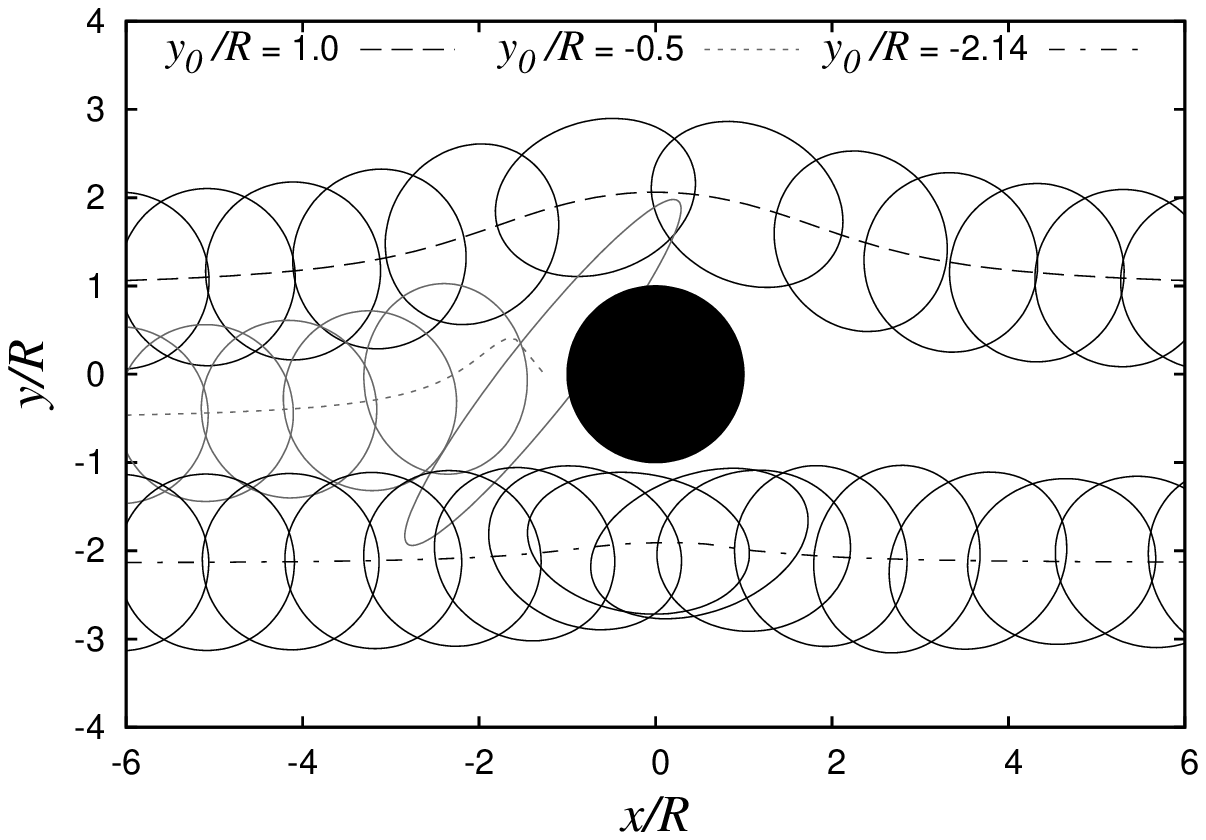}
 \caption{Motion of initially circular vortex patches past an island of the same size ($R = a$) calculated using the elliptic model. The irrotational background flow has strength $U/(\omega R) = 0.1$. The motion of vortex patches is shown for three different values of the release height $y_0$. Each motion is visualised by the path of the ellipse centroid alongside snapshots of the elliptic shape at time intervals of $10/\omega$.  }
 \label{fig:island_model}
\end{figure}

Under the elliptic model, we can investigate the deformation of initially circular vortex patches released at a height $y_0$ far upstream of the island ($x_0 \ll -a$). Depending on the strength of the background flow and the release height, the patches undergo one of three different motions: they either pass above the island, pass below the island, or pass sufficiently close to the island that the patch is indefinitely extended.  Sufficiently close to the point vortex separatrix, another family of solutions may exist: vortex patches that loop around the island a number of times before continuing downstream. However, we were unable to find such solutions numerically, suggesting that any such solutions are limited to a very small range of release heights $y_0$. Examples of the three different motions that were observed are shown in Figure~\ref{fig:island_model}. 

\begin{figure}
 \centering
 \includegraphics[width=0.7\textwidth]{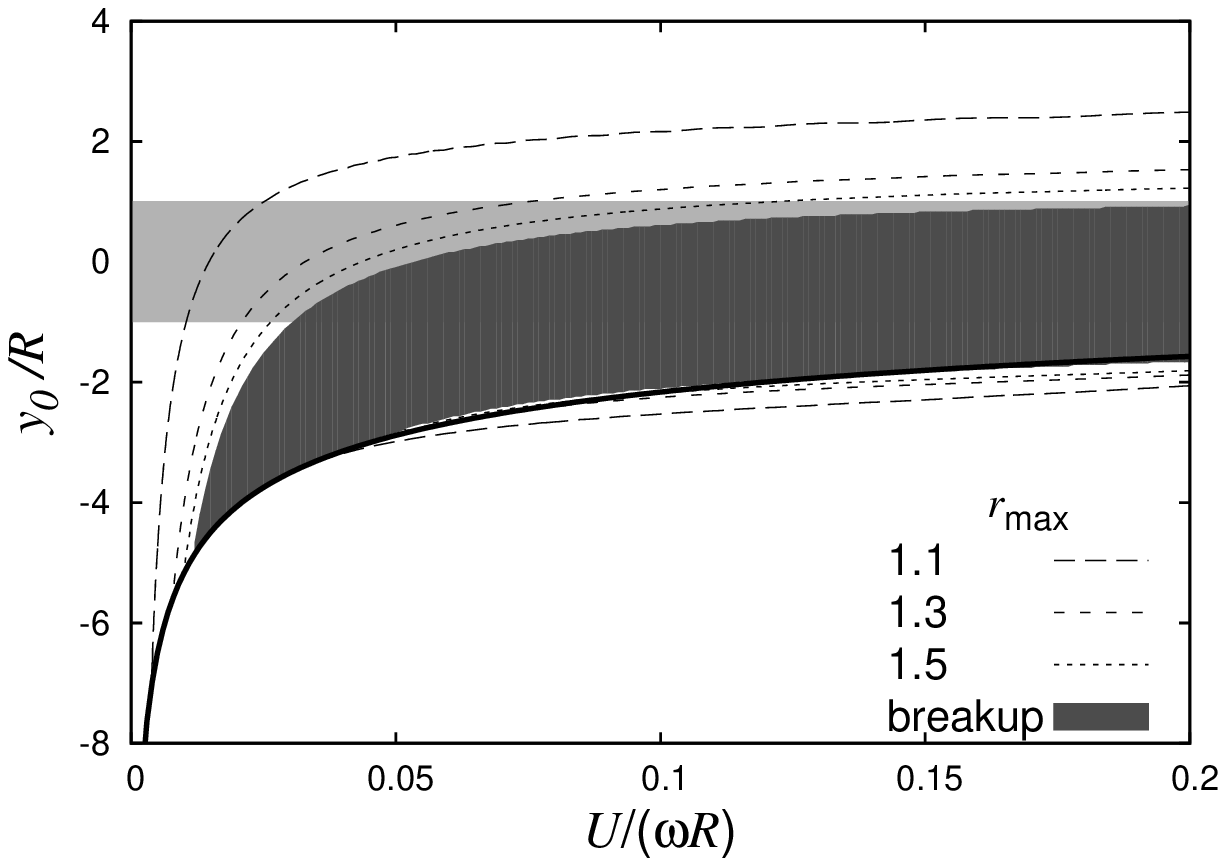}
 \caption{Maximum deformation $r_{\rm max}$ of elliptic vortex patches with the same size as the island ($R = a$) over a range of background flow strengths $U$ and release heights $y_0$.  The region of parameter space for which vortex patches are indefinitely extended is shaded in dark grey. Elsewhere, contours of $r_{\rm max}$ are shown. For further information, the location of the separatrix for point vortices is shown (heavy line) as is the location of the island (light grey shaded region).}
 \label{fig:island_flow_distortion}
\end{figure}

The deformation of an initially circular vortex patch can be investigated quantitatively by calculating the maximum deformation $r_{\rm max}$ as the patch passes the island. For a given ratio of island radius to patch radius ($a/R$), the elliptic model can then be used to quickly sweep over a range of $(U,y_0)$ parameter space.  Such an analysis is shown in Figure~\ref{fig:island_flow_distortion} for the case $a = R$. Other island sizes give qualitatively similar results.

The deformation of a patch depends significantly on whether it passes above or below the island.  This is a consequence of the location of the point vortex separatrix (Figure~\ref{fig:island_flow}b), which passes closer to the top of the island than it does the bottom (for $U/\omega > 0$). The path of a point vortex (of appropriate strength) gives a leading-order approximation to the path of the centre of the elliptic vortex patch. Thus patches starting just above the point vortex separatrix will pass closer to the island, and be deformed more, than those starting just below.  The smaller the background flow speed, the further the point vortex separatrix lies from the island. Consequently, for small background flow speeds ($|U| \ll |\omega| R$), there is very little deformation of the vortex patch, independent of release height. Meanwhile, for large background flow speeds ($|U| \gg |\omega| R$), the vorticity of the patch has very little effect on the motion and the centres of the vortex patches follow paths closer to those shown in Figure~\ref{fig:island_flow}(a). In this `ballistic' regime, those vortex patches that get deformed significantly correspond to vortex patches whose impact height is in, or close to, the range of impact heights covered by the island.

\subsection{Comparison with full solution}

\begin{figure}
 \centering
 \includegraphics[width=0.95\textwidth]{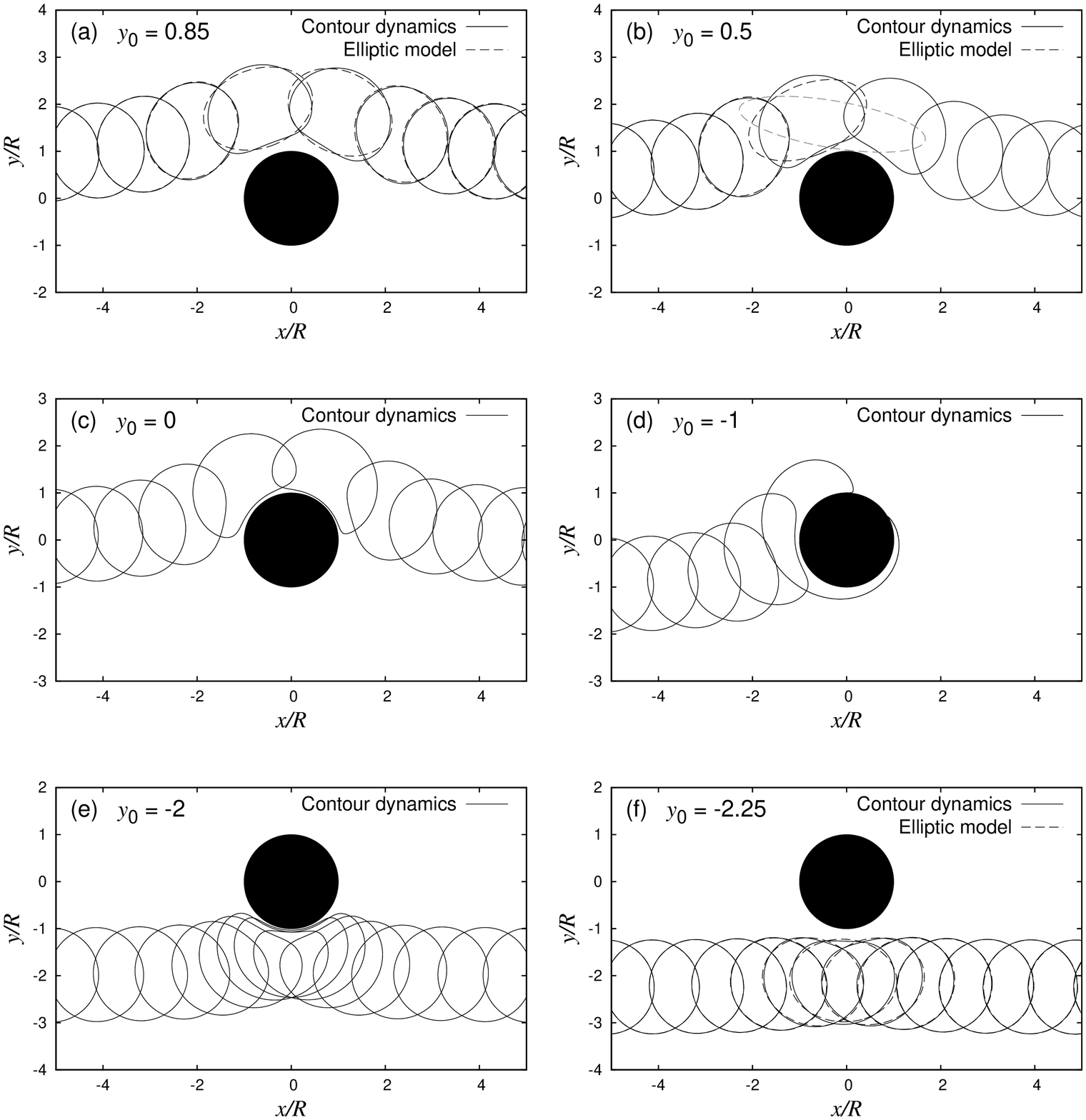}
 \caption{Motion of vortex patches with varying release heights driven towards an island by background flow; each panel shows a lower release height than the one before. The island and the patch have the same area, and the background flow has strength $U/(\omega R) = 0.1$. Solutions were calculated using the elliptic model (dashed lines) and using contour dynamics (solid lines). The patch motion is visualised by snapshots of the patch position at intervals of $10/\omega$. In (a), the patch passes above the island and the elliptic model ($r_{\rm max} = 1.53$) is in good agreement with the full solution; there is similar agreement in (f) when the patch passes below the island ($r_{\rm max} = 1.30$). For intermediate release heights (b--e), the elliptic model predicts indefinite extension, as shown in (b). For these intermediate heights, the patch is either deformed into a non-elliptical shape while passing the island before returning to an approximately circular shape downstream (b,c,e); or the patch is extended around both sides of the island (d).}
 \label{fig:island_U01}
\end{figure}

The elliptic model is useful in allowing us to rapidly investigate large regions of ($y_0,U$) parameter space, but it is only an approximation to the full solution.  For vortex patches that pass very close to the island, such as those that the elliptic model predicts are indefinitely extended, further investigation via calculation of the full solution is needed.

The full solution, calculated for selected values of $y_0$ with $U/(\omega R) = 0.1$ and $a = R$, is shown in Figure~\ref{fig:island_U01}.  From this figure, we note that the elliptic model is in good agreement with the full solution when $r_{\rm max} \lesssim 1.5$. For those values of $y_0$ for which patches are indefinitely extended under the elliptic model ($-2.11R < y_0 < 0.61R$ for this choice of parameters), the full solutions shown in Figure~\ref{fig:island_U01} reveal two possible behaviours: patches are either distorted into a non-elliptical shape, before returning to an approximately circular shape downstream; or, for an intermediate range of release heights, they are broken up as they get extended around both the top and bottom of the island.


\section{Conclusion}

We have investigated the deformation of two-dimensional patches of uniform vorticity in the vicinity of fluid boundaries both via the Hamiltonian elliptic model of Melander, Zabusky and Styczek, and via contour dynamics simulations. For a straight boundary, the distance of the vortex patch centroid from the boundary is fixed and the motion is integrable under the elliptic model. In fact, the problem reduces to the well known problem of an elliptic vortex patch in a constant straining flow as analysed by Kida \cite{Kida}. The elliptic model is a good approximation to the full solution even when the separation of the elliptic patch from the boundary is comparable to the dimensions of the patch. As the patch is moved away from the boundary, the straining flow generated by its image falls off as the inverse square of its distance from the boundary. Consequently, there is a rapid decrease in the deformation of the patch with distance from the boundary.

The motion of a vortex patch in a quarter plane demonstrates nonintegrable behaviour under the elliptic model. However, due to a separation in time scales between the time scale on which the patch rotates and the time scale on which the straining flow experienced by the patch varies (the former being shorter than the latter), there is an adiabatic invariant. The existence of an adiabatic invariant acts to limit the  deformation of vortex patches by preventing the accumulation of small deformations over time. We expect the motion of a vortex patch along boundaries with more complicated shapes to be similarly limited by the presence of an adiabatic invariant. When a background straining flow is added to the quarter plane geometry, vortex patches can become trapped in the corner region. Poincar\'{e} sections of vortex patches in this system, evolving under the elliptic model, reveal the model's chaotic nature.

We also investigated the deformation of a vortex patch in the vicinity of a circular island both with and without a background flow. Without a background flow, the motion of a vortex patch around a circular island is integrable under the elliptic model. The amount of deformation caused by an island is less than that caused by a straight boundary to a vortex patch at a similar separation, and increasingly so as the island radius is decreased. The amount of deformation also falls off more rapidly with distance from the island: the straining flow due to an island falls off asymptotically as the inverse fourth power of the separation between the patch and the island compared to an inverse square fall-off for a straight boundary. When a background flow is added, significant deformation of vortex patches is possible as they are driven towards the island boundary. The addition of an irrotational background flow that tends to uniform flow away from the island leads to three possible evolutions of vortex patches under the elliptic model: as they are advected towards the island, vortex patches either pass above the island, pass below the island, or are indefinitely extended. For those vortex patches that the elliptic model predicts are indefinitely extended, a closer examination using the full contour dynamics solution reveals two possible behaviours: either vortex patches are deformed into a strongly non-elliptical shape before passing around one side of the island and returning to an approximately circular shape downstream; or part of the vortex patch is split as part passes above the island and part passes below.

 \appendix

\section{Contour dynamics for a circular island}
\label{sec:island_CD}

Since the image vorticity distribution for a uniform vorticity patch outside a circular island has non-uniform vorticity, the derivation of the contour dynamics method for such a domain is more complicated. In fact, it is possible to apply the contour dynamics method to any finitely connected planar domain, as shown by Crowdy and Surana \cite{CD_complex_domains}, but here we just give a brief derivation of the result for a circular island.

The streamfunction at $\mathbf{x}$ due to both the self vorticity and the image vorticity of a vortex patch of vorticity $\omega$ occupying an area $A$ outside an island of radius $a$ is given by
\begin{equation}
 \psi(\mathbf{x}) = -\frac{\omega}{2 \pi} \int_A \ln \left( \frac{|\mathbf{x}-\mathbf{y}||\mathbf{x}|}{\left|\mathbf{x}-\mathbf{y}\frac{a^2}{|\mathbf{y}|^2}\right|} \right) \,d^2\mathbf{y}.
\end{equation}
The velocity is given by derivatives of the stream function, and the contour dynamics method consists of converting the resulting area integrals into integrals about the vortex patch boundary. Taking a spatial derivative, and splitting into to contributions from the self and image vorticity gives
\begin{equation}
 \nabla_\mathbf{x} \psi(\mathbf{x}) = -\frac{\omega}{2 \pi} \nabla_\mathbf{x} \int_A \ln \left( |\mathbf{x}-\mathbf{y}|\right) \,d^2\mathbf{y} - \frac{\omega}{2 \pi} \nabla_\mathbf{x} \int_A \ln \left( \frac{|\mathbf{x}|}{\left|\mathbf{x}-\mathbf{y}\frac{a^2}{|\mathbf{y}|^2}\right|} \right) \,d^2\mathbf{y}.
\end{equation}
In order to rewrite the second term, which represents the image vorticity, as a boundary integral, we rearrange it as
\begin{equation}
 \nabla_\mathbf{x} \psi(\mathbf{x}) = -\frac{\omega}{2 \pi} \nabla_\mathbf{x} \int_A \ln \left( |\mathbf{x}-\mathbf{y}|\right) \,d^2\mathbf{y} - \frac{\omega}{2 \pi} \nabla_\mathbf{x} \int_A \ln \left( \frac{|\mathbf{y}|}{\left|\mathbf{y}-\mathbf{x}\frac{a^2}{|\mathbf{x}|^2}\right|} \right) \,d^2\mathbf{y}.
\end{equation}
This allows both derivatives with respect to $\mathbf{x}$ to be rewritten as derivatives with respect to the dummy variable $\mathbf{y}$
\begin{equation}
 \nabla_\mathbf{x} \psi(\mathbf{x}) = \frac{\omega}{2 \pi} \int_A \nabla_\mathbf{y} \ln \left( |\mathbf{x}-\mathbf{y}|\right) \,d^2\mathbf{y} - \frac{\omega}{2 \pi}  \nabla_\mathbf{x} \left( \mathbf{x}\frac{a^2}{|\mathbf{x}|^2} \right) \mathbf{\cdot} \int_A \nabla_\mathbf{y} \ln \left( \left|\mathbf{y}-\mathbf{x}\frac{a^2}{|\mathbf{x}|^2}\right| \right) \,d^2\mathbf{y}.
\end{equation}
Then, applying the divergence theorem, these area integrals can be rewritten as integrals around the boundary of the vortex patch
\begin{equation}
 \nabla_\mathbf{x} \psi(\mathbf{x}) = \frac{\omega}{2 \pi} \int_{\partial A} \ln \left( |\mathbf{x}-\mathbf{y}|\right) \,d\mathbf{n} - \frac{\omega}{2 \pi} \frac{a^2}{|\mathbf{x}|^2} \left( \mathbf{I}-2\frac{\mathbf{x}\mathbf{x}}{|\mathbf{x}|^2} \right) \mathbf{\cdot} \int_{\partial A} \ln \left( \left|\mathbf{y}-\mathbf{x}\frac{a^2}{|\mathbf{x}|^2}\right| \right) \,d\mathbf{n},
\end{equation}
which is the desired contour dynamics formulation.

\begin{acknowledgments}
The authors would like to thank the Woods Hole Oceanographic Institute whose Geophysical Fluid Dynamics program provided the inspiration and basis for the preceding research. The authors remain grateful to Prof.\ David Dritschel for providing a copy of his contour dynamics code on which our code is based. AC is grateful for support from an EPSRC studentship. PJM was supported by U.S.~Dept.\ of Energy Contract \# DE-FG05-80ET-53088.
\end{acknowledgments}


%

\end{document}